\documentclass{amsart}
\usepackage[utf8]{inputenc}
\usepackage{amsfonts}
\usepackage{amsmath}
\usepackage{amssymb}
\usepackage{mathrsfs} 
\usepackage{enumerate}
\usepackage{xcolor}
\usepackage{rotating}
\usepackage[justification=centering]{caption}
\usepackage{hyperref}

\usepackage{scalerel,stackengine}
\stackMath
\newcommand\rwhat[1]{%
\savestack{\tmpbox}{\stretchto{%
  \scaleto{%
    \scalerel*[\widthof{\ensuremath{#1}}]{\kern-.7pt\bigwedge\kern-.7pt}%
    {\rule[-\textheight/2]{1ex}{\textheight}}
  }{\textheight}%
}{0.5ex}}%
\stackon[1pt]{#1}{\tmpbox}%
}
\newcommand{\fuzzy}{%
           \mathrel{\raisebox{.1em}{%
           \reflectbox{\rotatebox[origin=c]{90}{\scalebox{1.2}[1.2]{=}}}}}}
\usepackage{stackengine}

\usepackage{tikz}
\usetikzlibrary{arrows.meta,automata,quotes}

\theoremstyle{plain}
\newtheorem{thm}{Theorem}
\newtheorem{cor}[thm]{Corollary}
\newtheorem{lem}[thm]{Lemma}
\newtheorem{prop}[thm]{Proposition}
\newtheorem{conj}[thm]{Conjecture}

\theoremstyle{definition}
\newtheorem{defi}[thm]{Definition}

\newtheorem{problem}[thm]{Problem}
\newtheorem{obs}[thm]{Observation}

\newcommand{\tb}{\ensuremath{\mathrm{TB}}}

\renewcommand{\L}{\mathrm L}
\newcommand{\R}{\mathrm R}

\newcommand{\nz}{\mathbb N_0}

\newcommand{\GL}{G^{\mathcal L}}
\newcommand{\GR}{G^{\mathcal R}}
\newcommand{\HL}{H^{\mathcal L}}
\newcommand{\HR}{H^{\mathcal R}}

\numberwithin{subcase}{case}
\numberwithin{subsubcase}{subcase}

\newcommand{\leac}{\mathrel{\mathsurround=0pt \mbox{$\lhd$\raisebox{1pt}{\tiny \textbf{\textbar }}}}}
\newcommand{\geac}{\mathrel{\mathsurround=0pt \mbox{\raisebox{1pt}{\tiny \textbf{\textbar }}$\rhd$}}}

\newcommand{\cg}[2]{\left\{ #1\!\mid \!#2\right\}}
\newcommand{\up}{\ensuremath{\uparrow\,}}
\newcommand{\down}{\ensuremath{\downarrow\,}}

\newcommand{\nil}{\varnothing}
\renewcommand{\tilde}{\widetilde}
\renewcommand{\bar}{\overline}

\newcommand{\B}{\mathcal B}
\renewcommand{\ge}{\geqslant}
\renewcommand{\le}{\leqslant}

\title{Constructive comparison in bidding combinatorial games}

\author{Prem Kant}
\address{Prem Kant, IIT Bombay, India}
\email{premkant@iitb.ac.in}
\author{Urban Larsson}
\address{Urban Larsson, IIT Bombay, India}
\email{larsson@iitb.ac.in}
\author{Ravi K. Rai}
\address{Ravi Kant Rai, University of Liverpool, United Kingdom}
\email{r.k.rai@liverpool.ac.uk}
\author{Akshay V. Upasany}
\address{Akshay Vilas Upasany, IIT Bombay, India}
\email{akshay.upasany@iitb.ac.in}

\date{\today}

\begin{document}

\begin{abstract}
A class of discrete Bidding Combinatorial Games that generalize alternating normal play was introduced by Kant, Larsson, Rai, and Upasany (2022). The major questions concerning optimal outcomes were resolved. By generalizing standard game comparison techniques from alternating normal play, we propose an algorithmic play-solution to the problem of game comparison for bidding games. 
We demonstrate some consequences of this result that generalize  classical results in alternating play (from Winning Ways 1982 and On Numbers and Games 1976). In particular, integers, dyadics and numbers have many nice properties, such as group structures, but on the other hand the game $*=\cg{0}{0}$ is non-invertible. We state a couple of thrilling conjectures and open problems for readers to dive into this promising path of bidding combinatorial games. 
\end{abstract}

\maketitle
\section{Introduction}
In a recent paper \cite{KLRU}, we introduce Discrete Bidding Combinatorial Games that generalize alternating normal play. All  essential structure problems regarding outcomes were resolved. In this paper we study the same families of games, by generalizing standard game comparison techniques from alternating normal play \cite{BCG2001,C1976, S2013}; we propose a  constructive i.e. algorithmic solution to the problem of game comparison. 

Our bidding is similar to Richman auctions; for further references see  \cite{larsson2019endgames,DP2010,LPR2021,LLPU1996}.  
Let us review the basic definitions and main results of our main reference \cite{KLRU}. 
A {\em game form} $G$ is recursively defined, with $G=\cg{\GL}{\GR}$, where $\GL$ is the set of all Left options and $\GR$ is the set of all Right options. If $\GL=\varnothing$ or $\GR=\varnothing$ then $G$ is Left terminal or Right terminal respectively, i.e., the current player, Left or Right, cannot move. In the case when  $\GL=\GR= \varnothing$, then $G=0$ is terminal irrespective of move order. A typical Left (Right) option of a game form $G$  is written  $G^L$ ($G^R$). Game forms are finite and contain no cycles, i.e. each game form has finitely many options, and the birthday (rank of game tree) is finite, implying that each play sequence is finite.\footnote{The formal birthday is the rank of the literal form game tree. Since we do not yet study reduction techniques of games, we will not distinguish between  birthday and formal birthday.} The game form $H$ is a \emph{subgame/follower}  of a game form $G$ if there exists a path of moves, perhaps empty, (in any order of play) from $G$ to $H$.

For the set of natural numbers $\mathbb{N} = \{1, 2, 3, \ldots \}$, define $\nz = \mathbb {N}\cup\{0\}$. Consider a {\em total budget} $\tb\in \nz$. A {\em bidding game} is a game form $G$ together with the total budget $\tb$ and is denoted by $(\tb, G)$. An instance of a bidding game is a triple $\left(\tb, G, \tilde p\,\right)$, where we take a note of Left's part of the budget, $\tilde p\in \B =\{0,\ldots, \tb,\rwhat 0,\ldots,  \rwhat \tb\}$. Here $\, \rwhat \cdot \,$ indicates that Left holds the tie-breaking marker. If $\tb$ is understood, we write $\left(G, \tilde p\,\right)$. The word ``game'' is used interchangeably where each surrounding context decides its precise meaning. 

There is a bidding round at a terminal position, and a player who wins a bid but cannot move, loses. 
  The player that does not hold the marker can {\em pass} in any such situation, by bidding zero.

Consider a non-terminal position $(G,\hat p)$ (Left holds the marker). There are four cases to establish the next game (in case there is one). An instance of bidding is an ordered pair, $(\ell,r)$, where Left bids $\ell$ and Right bids $r$.
\begin{itemize}
\item[(i)] $\left(G,\rwhat p\right)\rightarrow \left(G^L,\rwhat{ p-\ell}\right)$;   Left outbids Right, i.e. $\ell>r$.
\item[(ii)] $\left(G,\rwhat p\right)\rightarrow \left(G^L, p-\ell\right)$;  Left outbids Right, while including the marker,  $\ell \ge r$. 
\item[(iii)] $\left(G,\rwhat p\right)\rightarrow \left(G^L, p-\ell\right)$;  Left wins a tie, i.e. $\ell = r$. 
\item[(iv)] $\left(G,\rwhat p\right)\rightarrow \left(G^R, \rwhat{ p+r}\right)$;  Right outbids Left, i.e.  $\ell<r$.
\end{itemize}
Note that the third item is included in the second; but from a recreational play point of view,  those bids are distinguished.\footnote{If Left does not want to hold the marker in the next round, then she should of course include it to the current bid; if she prefers to keep the marker in case she wins the bid, she would act differently.}. 
Observe that in case of a tie, the marker shifts owner. This automatic rule is to the core of our generalization of alternating play. Namely $\tb=0$ corresponds to alternating normal play rules. 

\begin{thm}[First Fundamental Theorem, \cite{KLRU}]\label{thm:pure}
Consider the bidding convention where the tie-breaking marker may be included in a bid. For any game $\left(\tb, G,\tilde p\,\right)$ there is a pure strategy subgame perfect equilibrium, computed by standard backward induction.
\end{thm}
An important consequence of our auction definition (above item (ii)) is that ``the last move wins'' is equivalent with ``a player who cannot move loses''. A player who has the larger part of the budget, or half the budget together with the marker, is called the (currently) {\em dominating player}. 
 A player is {\em strictly dominating} if they have a strictly larger budget than their opponent.
\begin{cor}[Last Move Wins, \cite{KLRU}]\label{lem:lastplayer}
If a dominating player has an option from which the other player cannot move, then the dominating player wins the game.
\end{cor}

The pure subgame perfect equilibrium of a game $\left(\tb, G,\tilde p\,\right)$ is the {\em partial outcome}  $o\left(G,\tilde p\,\right)\in \{\R,\L\}$, where by convention the total order of the results is $\L>\R$, i.e. `Left wins' $>$ `Right wins'.  The {\em outcome} of the bidding game $(\tb, G)$ is $o(G)=o_\tb(G)$, defined via the $2(\tb+1)$ tuple of partial outcomes as $$o(G)=\left(o(G,\rwhat\tb),\ldots , o(G,\rwhat 0\, ),o(G, \tb),\ldots , o(G,0)\right).$$

\begin{defi}[Outcome Relation]\label{def:outrel} Consider a fixed $\tb$ and the set of all budgets $\B$. Then for any games $G$ and $H$, $o(G) \ge o(H)$ if,$~\forall~\tilde p \in \B, o(G, \tilde p) \ge o(H, \tilde p)$.\footnote{It is easy to check that the outcome relation is reflexive, antisymmetric and transitive. Hence the set of all outcomes together with this relation is a poset.}\end{defi}
Note that the outcomes inherit a partial order from the total order of the partial outcomes.


The next two results have good use in this paper, and together will be referred as ``MMW" (Monotonicity and Marker Worth).
\begin{lem}[Outcome Monotonicity, \cite{KLRU}]\label{thm:monotone}
Consider a fixed $\tilde p\in \B$, with $p<\tb$. Then, for all games G, $o\left (G,\tilde p\, \right )\le o\left (G,\tilde{p+1}\right)$.
\end{lem} 

\begin{lem}[Marker Worth, \cite{KLRU}]\label{thm:markerworth}
Consider $\tb\in \mathbb N_0$. Then, for all games G, $o(G,\rwhat p) \le o(G,p+1)$.
\end{lem}
The main result of paper \cite{KLRU} will not be used here, but it is worthwhile restating it. An outcome is {\em feasible} if it satisfies MMW.

\begin{thm}[\cite{KLRU}]
For any total budget, for each feasible outcome $\omega$, there is a game $G$ such that  $o(G)=\omega$. 
\end{thm}
In Section~\ref{sec:sums} we review the definitions of disjunctive sum and the partial order of games. The main result of this paper is a constructive approach for game comparison, which is discussed in Section~\ref{sec:constructive}; this is a generalisation of comparison of number games from Section~\ref{sec:casestudynumbers}. In Section~\ref{sec:nogroup} we show that, when $\tb>0$, the game $* = \cg{0}{0}$ does not have an inverse. Thus, the set of all bidding games is not a group structure if $\tb>0$. In Section~\ref{sec:comparison} we study the influence of invertibility in game comparison. The next parts of the paper, Sections~\ref{sec:numbers}, \ref{sec:integer}, \ref{sec:dyad} focus on gmaes that are numbers, integers and dyadic rationals respectively. In Section~\ref{sec:infinitesimal} we study infinitesimals and zero games, and in Section~\ref{sec:future} we provide some interesting conjectures and open problems. 
\section{Sum, partial order and  numbers}\label{sec:sums}

Let us recall the definitions of disjunctive sum of game forms and partial order of games \cite{BCG2001, S2013}. The disjunctive sum of the game forms $G$ and $H$ is defined recursively as: $$G+H=\cg{G+\HL,\GL+H}{G+\HR, \GR+H},$$ Where $G+\HL=\{G+H^L: H^L\in\HL\}$, in case $\HL\neq\varnothing$, and otherwise the set is not defined and omitted. 
The {\em conjugate} of a game form $G$ is the game form where players have swapped positions, and is recursively defined as $\bar G=\cg{\bar{\GR}\,}{\,\bar{\GL}}$. 


The {\em partial order} of games is defined as usual. Consider games $G,H$. Then $G\ge H$ if, for all games $X$, $o(G+X)\ge o(H+X)$. Game equality satisfies $G=H$ if $G\ge H$ and $H\ge G$. 
The games $G>H$ ($G$ is greater than $H$) if $G\ge H$ but $H\not \ge G$. The games $G$ and $H$ are confused if $G\not\ge H$ and $H\not\ge G$; this is denoted $G  \fuzzy H$. The games $G \leac H$ ($G$ is less than or confused with $H$) if either $G<H$ or G $\fuzzy$ H. The game $G$ is {\em invertible} if there exist a game $G'$ such that $G+G'=0$. 

A game form $G$ is a {\em number} if for all $G^L$ and for all $G^R$, $G^L<G<G^R$, and all options are numbers.\footnote{For alternating play the definition of game $G$ being number is $H^L<H^R$ for every subposition $H$ of $G$ and every $H^L$ and $H^R$. However the direct implication of this is $G^L<G<G^R$ \cite{S2013}.} We will later prove that all numbers are invertible, and the inverse is the conjugate form. And we will prove that numbers are closed under addition, so they are a subgroup of the monoid of all bidding games. Since we observe infinitely many invertible elements, we also have infinitely many $0$s. 

The empty game satisfies: for all $X$,  $o(\cg{\nil}{\nil}+X)=o(X)$. Namely, independent of play sequence, the final auction in a follower of $\cg{\nil}{\nil}+ X$ will appear at the game  $\cg{\nil}{\nil}+\cg{\nil}{\nil}=\cg{\nil}{\nil}$ (an idempotent). We already introduced the name $0=\cg{\nil}{\nil}$.  We argue that this is the natural name: if a game $H=0$, then we may omit it in a sum of games. that is, for all games $G$, if $H=0$, then $G+H=G$. In fact, the proof is immediate by definition (and does not depend on the specifics of the auction).

\begin{thm}\label{thm:zero}
If the game $H=0$, then, for any game $G$, $H+G=G$.
\end{thm}
\begin{proof}
We must demonstrate that for all $Y$, $o(G+H+Y)=o(G+Y)$. Since $H=0$, we have that for all $X$, $o(H+X)=o(X)$. Take $X=G+Y$.
\end{proof}

By uniqueness, we mean with respect to equivalence class / game value (as opposed to `game form'). 

\begin{thm}[Uniqueness of Identity]\label{thm:uniqueiden}
Consider games $G$ and $H$. Suppose $G$ has an inverse $G'$; i.e. $G+G'=0$. Then  $G+H = G$ implies $H=0$.
\end{thm}
\begin{proof}
    Let $H$ be such that $ G+H =G $. Then for all $(X,\tilde p\,)$, $o(G+H+X, \tilde p\,) = o(G+X, \tilde p\,)$. 
    Now, for any game $Y$, taking $X= G'+Y$, gives the result.
\end{proof}






\section{A case study for numbers}\label{sec:casestudynumbers}






Left and Right are going to play a intricate bidding game $X$. Just before start, Right offers Left to include a game $G$ to play in a disjunctive sum with $X$. Now the question is, should she accept this kind offer? To make any decision, she inquires whether the game $G$ is a number. If he affirms, then she plays the game $G$ without  budget or marker. If she wins it then she accepts the offer and otherwise she rejects.
However if he does not reveal anything about $G$, 
then she has to wait until the next section.

Our first result states that, Left weakly prefers a number game form before the neutral element if and only if she wins playing without the marker. 
\begin{thm}[Number Comparison]\label{thm:markerlosenumber}
Consider any total budget and a number game $G$. Then $G\ge 0$ if and only if $o(G,0)=\L$.
\end{thm}
\begin{proof}
By definition, $G\ge 0$ if and only if $~\forall X $ and $\forall \tilde p \in \B$, $o(G+X, \tilde p) \ge o(X, \tilde p)$. That is, $G\ge 0$ if and only if $~\forall X $ and $\forall \tilde p \in \B$, $o(X, \tilde p) = \L~~\implies~o(G+X, \tilde p) = \L$. \\

\noindent $\Rightarrow$: Take $X=0$, and $\tilde p = 0$. Then since $o(0,0) =\L$, so $G\ge 0$ implies $o(G,0) =\L$.\\

\noindent $\Leftarrow$: Let us assume  $o(G,0)=\L$ and $o(X, \tilde p) = \L$. We must prove  $o(G+X, \tilde p) = \L$.\\

\noindent \textbf{Case 1:}  $o(X, \rwhat p) = \L$.

Suppose Left wins  $(X,\rwhat p)$ by bidding $\tilde \ell$.  We claim that Left wins $(G+ X, \rwhat p)$, by bidding the same $\tilde \ell$. 

To prove this, suppose first Right bids  $r<\ell$. In this case if Left's optimal bid in  $(X,\rwhat p)$ is $\rwhat \ell$, then she plays to $(G + X^L,  p-\ell)$, and if it is $\ell$, then she plays to  $(G + X^L, \rwhat{ p-\ell})$. There is a Left option $X^L$, in $(X,\rwhat p)$ since Right can bid $0$. Left wins $(G + X^L, \rwhat{ p-\ell})$ by induction, and she wins $(G + X^L,  p-\ell)$ by applying induction, using Case 2.

Now, if Right bids $r=\ell$, then Left wins the bid and plays to $(G + X^L,  p-\ell)$. Left wins this by applying induction using Case 2.

 Next assume that Right outbids Left with $r>\ell$. He plays to either $(G^R + X,\rwhat{p+r})$ for some $G^R \in \GR$ or to $(G + X^R,\rwhat{p+r})$ for some $X^R \in X^\mathcal{R}$; otherwise we are done. Using induction Left wins $(G + X^R,\rwhat{p+r})$  because $\forall X^R \in X^{\mathcal{R}}$ Left wins $(X^R,\rwhat{p+r})$. However for $(G^R + X,\rwhat{p+r})$, since  G is a number, $G^R > G$. Thus $o(G, 0) = \L \implies o(G^R, 0) = \L$, which implies $o(G^R + X,\rwhat{p+r}) = \L$, by induction and budget  monotonicity. \\

\noindent \textbf{Case 2:} $o(X,  p) = \L$. 

Suppose Left wins $(X,  p)$  by bidding $ \ell$. We claim that she wins $(G+ X,  p)$, by bidding the same $\ell$. 

To prove this, suppose first Right bids $r <\ell$ in game $(G+ X,  p)$. Left wins the bid and plays to $(G + X^L, p-\ell)$. There is a Left option $X^L$, since by assumption Left wins $(X,  p)$  by bidding $ \ell$; the case $\ell = 0$ will not arise because $r < \ell$.   She wins $(G + X^L,  p-\ell)$ using induction.

Suppose next Right bids  $r=\ell$. Then  he wins the bid and plays to  $(G^R + X,\rwhat{p+r})$ for some $G^R \in \GR$ or to $(G + X^R,\rwhat{p+r})$ for some $X^R \in X^\mathcal{R}$; otherwise we are done.  By using induction on Case 1, Left wins  $(G + X^R,\rwhat{p+r})$. For $(G^R + X,\rwhat{p+r})$, we do the analysis for one more step in which Left bids $\rwhat \ell$. Then for Right's bid $r_2$ the following subcases arise:

     \begin{enumerate}[i)]
       \item If $ r_2<\ell$, then Left wins the bid and plays to $(G^R + X^L, p+r-\ell)$, i.e. $(G^R + X^L, p) $.  From first paragraph of this case  Left wins $(G + X^L, p-\ell)$. Using budget monotonicity and property of number game  $G$, Left wins  $(G^R + X^L, p)$.
       \item If $ r_2=\ell$ then Left wins the bid and plays to $(G^{RL} + X, p)$. There is a Left option $G^{RL}$, since by assumption she wins $(G,0)$. So, after Right winning the first bid and playing to $(G^R, \rwhat 0)$, she has a defence, by playing to some $G^{RL}$. Here, Left wins   $(G^{RL} + X, p)$ using induction. 
        \item If $r_2>\ell$ then Right wins the bid and plays to either $(G^R + X^R,\rwhat{p+r + r_2})$ or $(G^{RR}+ X,\rwhat{p+r + r_2})$. For $(G^R + X^R,\rwhat{p+r + r_2})$, using the property of number game $G$ we have $o(G^R + X^R,\rwhat{p+r + r_2}) \ge o(G + X^R,\rwhat{p+r + r_2})$. Thus  Left wins $(G^R + X^R,\rwhat{p+r + r_2})$ using budget monotonicty and induction. Next using the property of number game $G^R$ we have  $o(G^{RR}+ X,\rwhat{p+r + r_2}) \ge o(G^{R }+ X,\rwhat{p+r + r_2})$. Note that $r_2>0$.  Thus, by repeating the analysis eventually we reach case i) or case ii).
    \end{enumerate}

Next assume that Right bids  $\tilde r$ with $r>\ell$. He wins the bid and plays to  either $(G^R + X,\rwhat{p+r})$ or $(G^R + X,p+r)$, for some $G^R \in \GR$, or to $(G + X^R,\rwhat{p+r})$ or $(G + X^R,p+r)$, for some $X^R \in X^\mathcal{R}$. From the analysis of the previous paragraph Left wins  $(G + X^R,\rwhat{p+r})$  and $(G^R + X,\rwhat{p+r})$. Further she wins $(G + X^R,p+r)$ by using induction. And for $(G^R + X,p+r)$, since $G$ is a number, $G^R > G$. Thus $o(G,0)=\L \implies o(G^R,0)=\L$, which implies $o(G^R + X,p+r) = \L$, by induction and budget monotonicity.
\end{proof}

Given a game $G$, Theorem~\ref{thm:markerlosenumber} is applicable only if it is a number. 
And verifying that 
itself involves checking $H^L<H<H^R$ for every subposition $H$ of $G$. 
 Furthermore, even if we are provided with a number game $G$, the computation of $o(G, 0)$ remains a necessary step. To overcome these challenges, we will next present a more efficient approach.

\section{A constructive main theorem}\label{sec:constructive}
In alternating play, it is well known and easy to prove that if Left wins both $G$ and $H$ playing second, then she wins $G+H$ playing second. If she wins $G$ playing second and $H$ playing first, then she wins $G+H$ playing first. The following lemma generalizes these results, to a particular bidding setting.

Consider a game $(G,\tilde p\, )$. Left plays a {\em 0-bid strategy} if she bids $0$ at each follower of $G$. Left has an {\em optimal 0-bid strategy}  if the 0-bid is optimal at each follower of $G$. Note that if a player has an optimal 0-bid strategy, they are not necessarily winning.


\begin{lem}[Additive Property]\label{lem:add}
Consider any total budget $\tb$. Suppose, for fixed marker owner $\, \tilde\cdot \,$, $o(G,p)=o(H,\tilde q\, )=\L$. If Left has an optimal 0-bid strategy in $(G,p)$ and $p+q\le \tb$, then $o\left (G+H,\tilde{p + q}\right)=\L$. 
Suppose $o(G,\rwhat{p}\,)=o(H,q)=\L$. If Left has an optimal 0-bid strategy in $\left(G,\rwhat{p}\,\right)$ and $p+q\le \tb$, then $o\left(G+H,\rwhat{p + q}\right)=\L$.
\end{lem}
\begin{proof}

%
We must prove, 
\begin{enumerate}[(i)]
    \item if $o(G,p)=o(H, q)=\L$, then $o(G+H, p + q)=\L$;
    \item if $o(G,p)=o\left(H,\rwhat q\, \right)=\L$, then $o\left(G+H,\rwhat{p + q}\right)=\L$;
    \item if $o(G,\rwhat p)=o(H,q)=\L$, then $o\left(G+H,\rwhat{p + q}\right)=\L$;
\end{enumerate}
where, in each case, we assume that Left has an optimal 0-bid strategy in $(G,\tilde p)$, and $p+q\le \tb$. 
The proof is by induction.\\ 
\noindent Case (i): Let $\ell$ be an  optimal Left bid in $(H,q)$. If $\ell>0$, then there exists a Left option in $H$ (since Right can pass). If $\ell=0$ then Right will win the bid (and a Left option in $H$ might not exist).

Suppose that Left bids $\ell$ in $(G+H,p + q)$. If Left wins the bid, she plays to $(G+H^L,p+ q-\ell)$, and wins by induction. 

If Right ties Left's $\ell$-bid, then he plays to $$\left(G^R+H, \rwhat{p+q+\ell}\right) {\text or } \left(G+H^R, \rwhat{p+q+\ell}\right)$$ (note that in this case $p+q\le \tb-\ell$). But, by monotonicity, since Left has an optimal 0-bid in $G$, then $o\Big(G^R,\rwhat{p+\ell}\,\Big)=o(H,q)=\L$. 
Also since $\ell$ is an optimal Left bid in $H$, we get $o(G,p)=o\Big(H^R,\rwhat{q+\ell}\,\Big)=\L$. Hence by induction, using (iii) and (ii), 
$$o\Big (G^R+H, \rwhat{p+q+\ell}\,\Big ) = o\Big (G+H^R, \rwhat{p+q+\ell}\,\Big)=\L.$$ 
If Right outbids Left by not using the marker, then he plays to $(G^R+H, p+q+\ell+1)$ or $(G+H^R, p+q+\ell+1)$ (note that in this case $p+q\le \tb-\ell-1$). In this case using Marker Worth, Right is not better than the previous situation where he ties the bid. 

Altogether $o(G+H, p+q)=\L$.\\

\noindent Case (ii). Let $\tilde{\ell}$ be an optimal Left  bid in $(H,\rwhat q)$. A winning Left option $H^L$ exists, because she holds the marker. 
 
 She bids the same in the game $(G+H,\rwhat{p+q})$. If she wins the bid and  $\tilde\ell=\rwhat\ell$ she plays to $(G+H^L, p+ q-\ell)$ and otherwise she plays to $\Big (G+H^L, \rwhat{p + q-\ell}\, \Big)$. In the first case, she wins by induction, since $o(G, p)=o(H^L, q-\ell)=\L$. In the second case, she wins by induction, since $o(G, p)=o(H^L, \rwhat{q-\ell})=\L$. 

If Right outbids Left, by monotonicity $r=\ell+1$, he plays to $\Big (G^R+H, \rwhat{p+q+\ell+1}\, \Big )$ or $\Big(G+H^R,\rwhat{p+q+\ell+1}\, \Big)$. But, since Left has an optimal 0-bid in $(G,p)$ by monotonicity,   $o(G^R,p+\ell+1)=o(H,\rwhat{q}\,)=\L$. And moreover, Left has a defence to a Right move in $H$, so  $o(G,p)=o\Big(H^R,\rwhat{q+\ell+1}\,\Big )=\L$. Hence, by induction, $$o\Big(G^R+H, \rwhat{p+q+\ell+1}\,\Big)=o\Big (G+H^R, \rwhat{p+q+\ell+1}\,\Big)=\L$$

Altogether $o(G+H,\rwhat{p+q})=\L$.\\

\noindent Case (iii). Let $\ell$ be Left's optimal bid in $(H,q)$. She bids the same in $(G+H,\rwhat{p + q})$ and if she wins the bid, she plays to $\left(G+H^L,\rwhat{p + q-\ell}\right)$ and wins by induction. Note that $H^L$ exists if $\ell>0$, and otherwise she cannot win the bid in $H$, because Right owns the marker. But in $\left(G+H,\rwhat{p+q}\right)$ she might win the bid if $\ell=0$ and Right passes. In this case, she has a winning move in $(G,\rwhat p\, )$ to $(G^L, p)$, and so, by using (i), by induction $o(G^L+H,p+q)=\L$.

Suppose Right wins the bid. Since Left owns the marker, we assume by monotonicity that he bids $r=\ell+1$, and plays to $$\Big (G^R+H,\rwhat{p+q+\ell +1} \Big ) \text{ or } \Big (G+H^R,\rwhat{p+q+\ell+1} \Big ).$$ Observe that $o\left (G^R,\rwhat{p+1}\right )=o\left (H,q\right )=\L$. Therefore induction together with monotonicity gives $o\Big(G^R+H,\rwhat{p+q+\ell +1} \Big)=\L$. And $o\left(G+H^R,\rwhat{p+q+\ell+1} \right)=\L$ by induction, since by assumption $o\left (G,\rwhat{p})=o(H^R,q+\ell+1\right )=\L$.

Altogether $o\left(G+H,\rwhat{p+q}\right)=\L$.
\end{proof}

One of the most celebrated results in alternating play theory states that Left wins $G$ playing second if and only if $G\ge 0$. We generalize this constructive-, algorithmic-, recursive-, play-, local-comparison to bidding play.\footnote{Constructive comparison  appears under many names in the literature.} For constructive comparison it is required to satisfy another recursive test: ``Does Left have an optimal $0$-bid strategy?''

\begin{thm}[Main Theorem]\label{thm:markerlose}
Consider a game form $G$ and any total budget. Suppose that Left has an optimal $0$-bid strategy in $(G,0)$. Then $G\ge 0$ if and only if $o(G,0)=\L$.
\end{thm}
\begin{proof}
By definition, $G\ge 0$ if for all $X$ and for all $\tilde p$, $o(X,\tilde p\, ) =\L$ implies  $o(G+X,\tilde p\, ) =\L$.\\ 

\noindent $\Rightarrow$: Assume $G\ge 0$. Take $X=0$ and $\tilde p = 0$. By $o(0,0) =\L$, the implication gives $o(G,0) =\L$.\\ 

\noindent $\Leftarrow$: 
          
\noindent {\bf Case 1:} Suppose $o(G,0)=\L$ and $o(X, p)=\L$, where Left has an optimal $0$-bid strategy in $(G,0)$. We must show that $o(G+X, p) = \L$. 

Consider a Left optimal bid in $(X,p)$, say  $\ell$. Assume that she bids $\ell$ in $(G+X,p)$.

If she wins the bid, she has a move to $(G+X^L,p-\ell)$, which she wins by induction. 

If Right ties the bid and plays in the $X$-component, Left wins by induction, using Case~2 below. Hence assume he ties and plays to 
\begin{align}\label{eq:GRX}
\left (G^R+X,\rwhat{p+\ell}\, \right ). 
\end{align}

Now, since by assumption, Left has an optimal 0-bid strategy in $G^R$, Lemma~\ref{lem:add} applies; since $o(G^R,\rwhat{0}\, )=o(X,p)=\L$, by monotonicity  Left wins \eqref{eq:GRX}. 

If Right outbids Left and includes the marker, then Monotonicity implies that his result cannot be better than in the previous paragraph; if he does not include the marker, then Marker Worth implies the same.\\

\noindent{\bf Case 2:} Suppose next that $o(G, 0) = \L$ and $o(X, \rwhat{p})=\L$, where Left has an optimal 0-bid strategy in $(G,0)$.  We must show that $o(G+X, \rwhat{p} ) = \L$.  Note that by the assumption $o(X, \rwhat{p} )=\L$, Left has a (winning) move in $X$ (Right can bid 0 and force Left to move).

Consider an optimal Left bid $\tilde \ell$ in $(X,\rwhat p\,)$, and assume that Left bids $\tilde \ell$ in $(G+X,\rwhat{p}\,)$. If she wins the bid by including the marker, she has a move to $(G+X^L,p-\ell\,)$, and otherwise she has a move to $(G+X^L,\rwhat{p-\ell}\,)$. In both cases Left wins by induction,  
since, by assumption, $o(X^L, p-\ell )=\L$ or $o(X^L, \rwhat{p-\ell}\, )=\L$. 
Similarly, if Right outbids Left and plays in the $X$-component, Left wins by induction. 

Hence suppose Right outbids Left and plays to 
\begin{align}\label{eq:GRX1}
\Big (G^R+X,\rwhat{p+\ell+1}\, \Big ). 
\end{align}
 Left bids $\ell$, and, unless Right outbids or ties Left's bid, she plays in the $G$-component to $(G^{RL}+X,\rwhat{p+1}\, )$ such that $o(G^{RL},0)=\L$. She wins by induction, since she has an optimal 0-bid strategy in $G^{RL}$   and by Monotonicity. 
 Assume that Right outbids Left and plays to 
 \begin{align}\label{eq:GRXR}
 \Big (G^{R}+X^R,\rwhat{p+2\ell+2}\, \Big ). 
 \end{align}
 Then we use Marker Worth together with Lemma~\ref{lem:add}. Namely $o(G^R,\rwhat{0}\, )=\L$ implies $o(G^R,1)=\L$, and Left has an optimal 0-bid strategy in $G^R$. Combine this with $o\Big(X^R,\rwhat{p+\ell+1}\,\Big)=\L$, to see that Left wins \eqref{eq:GRXR} (by using also Monotonicity).
 
 If Right ties Left's bid then she wins the bid by giving away the marker and playing to $(G^R + X^L, p+1)$. We have  $o(G^R,1)=\L$ with Left's optimal 0-bid strategy in $G^R$. Also,  since $o(X, \rwhat p) = \L$, we have $o(X^L, p-\ell) = \L$. Hence, by using Monotonicity and Lemma~\ref{lem:add}, we get $o(G^R + X^L, p+1) = \L$. 
 
 If Right outbids Left and instead plays to  
 \begin{align}\label{eq:GRR}
     \Big (G^{RR}+X,\rwhat{p+2\ell+2}\,\Big ), 
 \end{align}
 then observe that, since Left has an optimal 0-bid strategy in $(G^R,\rwhat 0)$, then $o(G^{RR},\rwhat 1)=\L$. And so, by Marker Worth, $o(G^{RR},2)=o(X,\rwhat{p}\, )=\L$. Therefore, since Left has an optimal 0-bid strategy in  $G^{RR}$, Lemma~\ref{lem:add} together with Monotonicity implies  $$o\Big (G^{RR}+X,\rwhat{p+2\ell+2}\, \Big )=~\L .$$ 

\end{proof}
We summarize the result in terms of how it is used  in applications.
\begin{cor}[Constructive Comparison Tests]\label{cor:main}
Consider any bidding game $(\tb, G)$.
\begin{enumerate}
    \item If $o(G, 0) = \L$, with a Left 0-bid strategy, then $G\ge 0$.
    \item If $o(G, \rwhat{\tb}) = \R$, with a Right 0-bid strategy, then $G\le 0$.
    \item If $o(G, 0) = \L$, with a Left 0-bid strategy and  $o(G, \rwhat{\tb}) = \L$, then $G > 0$.
    \item If $o(G, \rwhat{\tb}) = \R$, with a Right 0-bid strategy, and $o(G, 0) = \R$ then $G<0$.
    \item If $o(G, 0) = \R$ and $o(G, \rwhat{\tb}) = \L$ then {\textnormal{$G\fuzzy 0$}}.
    \item If $o(G, \rwhat{\tb}) = \L$ and either $o(G, 0) = \R$ or $o(G, 0) = \L$, with a Left 0-bid strategy, then $G\geac 0$.
    \item If $o(G, 0) = \R$ and either $o(G, \rwhat{\tb}) = \L$ or $o(G, \rwhat{\tb}) = \R$,  with a Right 0-bid strategy, then $G\leac 0$.
\end{enumerate}
\end{cor}



\begin{proof}
This follows from Theorem~\ref{thm:markerlose}.
\end{proof}

The statement of the Main Theorem (Theorem~\ref{thm:markerlose}) raises some questions, especially with respect to the similar theorem for games that are numbers in Section~\ref{sec:casestudynumbers}.
\begin{problem}\label{prob:main}
Concerning the similarities of Theorems~\ref{thm:markerlosenumber} and~\ref{thm:markerlose}, some questions arise:
\begin{enumerate}[(i)]
    \item Is it true that ``Left has a  winning 0-bid strategy in $(G,0)$'' if and only if ``for every Right option $H^R$, $H$ a follower of $G$, there is an answer by Left, $H^{RL}$''?
    \item Is it true that the second property in (i) holds if and only if $G$ is a non-negative number?
    \item Is it true that if $G$ is a number then each player optimally plays a 0-bid strategy?
    \item May we put the Left-bidding-0 strategy inside the equivalence in Theorem~\ref{thm:markerlose}, i.e. is it true that ``$G\ge 0$ if and only if $o(G,0)=\L$ and Left has an optimal 0-bid strategy in $G$''?
\end{enumerate}
\end{problem}
Item (iv) does not seem to be valid, since the first implication in the proof does not give a $0$-bidding strategy. But the reverse direction holds. 
We observe that we may not remove the proviso of a Left winning $0$-bid strategy in the statement of Theorem~\ref{thm:markerlose}.

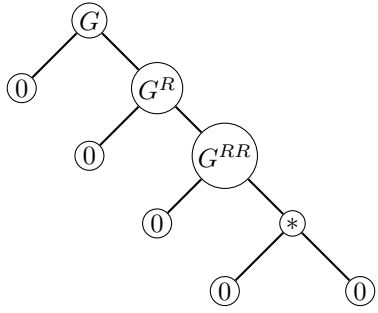
\begin{figure}[ht]
\begin{center}

\begin{tikzpicture}[scale = 0.9]
\begin{scope}[every node/.style={circle, fill=white,inner sep=1, draw}]  
    \node (0) at (0,0) {$G$};;
    \node (1) at (1,-1) {$G^R$};
    \node (2) at (2,-2) {$G^{RR}$};
    \node (3) at (3,-3) {$*$};
    \node (4) at (4,-4) {$0$};
    \node (5) at (-1,-1) {$0$};
    \node (6) at (0,-2) {$0$};
    \node (7) at (1,-3) {$0$};
    \node (8) at (2,-4) {$0$};
    \draw (0, 1) {};
\end{scope}
\begin{scope}[>={Stealth[black]},
              every edge/.style={draw=black,thick}] 
    \path [-] (1)edge(0);
    \path [-] (2)edge(1);
    \path [-] (3) edge (2);
    \path [-] (4) edge (3);
    \path [-] (5) edge (0);
    \path [-] (6) edge (1);
    \path [-] (7) edge (2);
    \path [-] (3) edge (8);
\end{scope}
\end{tikzpicture}
\caption{The game form $G$ in Observation~\ref{obs:nozero}, for $\tb\in \{ 4,5\}$.}
\label{fig:nonzero}
\end{center}
\end{figure}

\begin{obs}[No Generic $0$-optimal Strategy]\label{obs:nozero}

We demonstrate that, for any non-zero total budget there is a game form $G$ such that $o(G,0)=\L$, but Left does not optimally bid 0 at each follower. 
 Assume that Right knows that Left bids $0$ at each follower of some game $G$, that is yet to be constructed. Set $G^L=G^{RL}=\cdots = G^{R\cdots RL}=0$. He bids $\$1$ at each such round, for otherwise he loses. After $x>\tb/2$ rounds, the game has reached $(G^{R\cdots R},x)$. Let $H=G^{R\cdots R}=*$.  Left must outbid Right at this penultimate stage $H$, because ``last move wins". In optimal play she must bid $\$ x>0$ to win.  
For example, for $\tb=4$, the game $G=\cg{0}{\cg{0}{\;\up}}$ suffices, and is depicted in Figure~\ref{fig:nonzero}. Here the game $\up = \cg{0}{*}$. For $\tb=2$, we take instead the game $G=\cg{0}{\;\up}$. 
Observe that our game is not a number, since $*$ is a follower, and $*$ is not a number.
\end{obs}

Problem~\ref{prob:main} (i) and (ii) motivates (iii), for which we have a solution.

\begin{prop}[Zero-bids at Numbers]\label{prop:0bidnumber}
If $G$ is a number, then both optimally play 0-bid strategies.
\end{prop}
\begin{proof} Consider a number game $G$. If Left wins the bid, then she plays to $G^L<G$, which  does not improve her situation. And by Monotonicity she cannot gain by bidding $\ell>0$. Right faces the analogous sitation. 
 Since each follower of a number game $G$ is also a number, the result follows.
\end{proof}

Hence Theorem~\ref{thm:markerlosenumber} is  a consequence of Theorem~\ref{thm:markerlose}.




\section{No group structure}\label{sec:nogroup}
In alternating play, game families are groups. In bidding play games are closed with respect to addition, and there is a neutral element namely the empty game $0$. But there exist non-invertible elements when $\tb>0$.




 
 \begin{thm}[No Group Structure]\label{thm:noninv}
 Consider $\tb>0$. 
 There is no game $G$ such that $*+G=0$.
 \end{thm}
 \begin{proof}
 The proof is by way of contradiction. 
 The inequality $*+G\ge 0$ implies that Left wins $*+G$ if Right holds the marker (take $X=0$). This implies that there exists $G^L$ such that 
 \begin{align}\label{eq:noinv}
 o\left (G^L, 0\right )=\L .
 \end{align}
 Namely, Right begins in $*$, and after that Left plays her winning response in $G$; Left cannot win the first bid, Right can play in $*$; if Right passes the second bid, then Left gets to play and wins.
 
 On the other hand, she also wins if Right starts by playing in $*$, and then outbids Left and plays to $(G^R,\rwhat{1}\,)$. That is,  for any $G^R$, 
  \begin{align}\label{eq:noinv2}
 o\left (G^R,\rwhat{1} \right )=\L .
 \end{align}
 
 Observe, by symmetry, (with $*+G\le 0$) \eqref{eq:noinv} implies that there exists a $G^R$ 
 such that $o(G^R,\rwhat{\tb})=\R$. But then, by Monotonicity, for the same $G^R$,  $o(G^R,\rwhat{1}\,)=\R$, contradicting \eqref{eq:noinv2}.
\end{proof}

Thus, if $\tb>0$, then bidding games do not have a group structure. In general, we conjecture (Conjecture~\ref{conj:invconj}) that if a game has an inverse, then this game is its conjugate.

\section{Comparative Analysis}\label{sec:comparison}

In this section, we continue exploring comparison of bidding games. 
The existence of inverse becomes crucial.


\begin{obs}[Constructive Game Comparison]
Suppose that $H$ has an inverse, say $H'$. Then $G\ge H$ if Left has a winning 0-bid strategy in $(G+H',0)$.
\end{obs}

Games preserve order with respect to addition.

\begin{thm}[Order Preservation]\label{thm:orderpres}
Consider games $G,H$ and $J$. If  $J\ge H$, then $G+J\ge G+H$. 
\end{thm}
\begin{proof} This is immediate by definition of $J\ge H$. 
\end{proof}
For the converse of order preservation, the inverse is required.

\begin{thm} Consider games $G,H$ and $J$. Suppose $G$ has an inverse $G'$. Then $G+J\ge G+H$ implies $J\ge H$.
\end{thm}
\begin{proof}
    By $G+J\ge G+H$, for all $(X, \tilde p)$, $o(G+J+X, \tilde p\,) \ge o(G+H+X, \tilde p\,)$. Now, for any game $Y$, taking $X= G'+Y$, gives the result.
\end{proof}

Further, order preservation with respect to addition under strict inequality requires inverse.  
\begin{thm}\label{thm:strictorder} Consider games $G,H$ and $J$. Suppose $G$ has an inverse $G'$. Then $H<J$ if and only if  $G+H < G+J$.
\end{thm}

\begin{proof}
    Let us assume that $H<J$. Then there exists $(Y, \tilde q\,)$ such that  
\begin{align}
    o(H+Y, \tilde q\, ) &< o(J+Y,  \tilde q\,) \label{op2}
\end{align}
    Since $H<J$, we also have $H\le J$. Thus using Theorem \ref{thm:orderpres}, we have $G+H \le G+J$. 
    Next,  to prove $G+H<G+J$ we need to show there exists $(X,\tilde p\,)$ such that 
    $o(G+J+X, \tilde p\,) = \L~{\rm but}~ o(G+H+X, \tilde p\,) = \R$. 
    On the contrary, let us assume that there does not exist any such $(X,\tilde p\,)$. Then, for all $(X,\tilde p\,)$, 
$o(G+J+X, \tilde p\,) = o(G+H+X, \tilde p\,)$. 
By taking $X= G'+ Y$ and $\tilde p = \tilde q $ we get $o(J+ Y, \tilde q\,) = o(H+ Y, \tilde q\,)$.
This contradicts $\eqref{op2}$. Hence $G+H<G+J$. The other direction follows similarly.
\end{proof}

\begin{cor}\label{cor:inverseineq}
    Suppose a game $G$ has an inverse $G'$. Then $G>0$ if and only if $G'<0$.
\end{cor}
\begin{proof}
    This follows from Theorem~\ref{thm:strictorder}.
\end{proof}
The following theorem shows transitivity of order.
\begin{thm}\label{thm:ordertransitivity} Consider games $G,H$ and $J$. If $G\ge H$ and $H\ge J$ then $G\ge J$.
\end{thm}
\begin{proof}
    By $G\ge H$, for all $(X, \tilde p)$, $o(G+X, \tilde p) \ge o(H+X, \tilde p)$. Also since $H\ge J$, therefore, for all $(X, \tilde p)$, $o(H+X, \tilde p) \ge o(J+X, \tilde p)$. Thus, for all $(X, \tilde p)$, we have $o(G+X, \tilde p) \ge o(J+X, \tilde p)$. Hence $G\ge J$.
\end{proof}
Transitivity of order also holds with respect to strict inequality.
\begin{thm}\label{thm:strictordertransitivity} Consider games $G,H$ and $J$. If $G>H$ and $H>J$, then $G>J$.
\end{thm}
\begin{proof}
Suppose $G>H$ and $H>J$. This implies $G\ge H$ and $H \ge J$, respectively. By using Theorem~\ref{thm:ordertransitivity} we get $G \ge J$. Further
    since $G>H$, there exists a $(X, \tilde p)$ such that $o(G+X, \tilde p) = \L$ but  $o(H+X,\tilde p) = \R$. Since $H>J$, for the same $(X, \tilde p)$, we get that $o(H+X,\tilde p) = \R$ implies $o(J+X,\tilde p) = \R$. Hence there exist $(X, \tilde p)$ such that $o(G+X, \tilde p) = \L$ but $o(J+X,\tilde p) = \R$.
\end{proof}
In alternating play, if $G>0$, then Left wins irrespective of who starts the game. A similar result is also true for the bidding games.
\begin{thm}\label{thm:Ggr0}
    If the bidding game $G>0$, then for  any budget partition $\tilde p\in \B$,  $o(G, \tilde p) = \L$.
\end{thm}
\begin{proof}
    From Monotonicity it is sufficient to prove the following:
    \begin{enumerate}[i)]
        \item $o(G, 0) = \L$ and
        \item $o(G,\rwhat{0}) = \L$.
    \end{enumerate} 
     \textit{Proof of i). } Since $G>0$, thus $o(G,0) \ge o(0,0) = \L$.\\
     \noindent \textit{Proof of ii). } On the contrary let us assume that $o(G,\rwhat{0}) = \R$. Since $G>0$, there exists a $(X, \tilde p)$ such that 
     \begin{equation}\label{eq1}
        o(G+X, \tilde p) = \L ~{\rm but}~ o(X,\tilde p) = \R
     \end{equation}
     From Lemma~\ref{lem:add}, $o(G, \rwhat 0 ) = \R$ and $o(X, \tilde p) = \R$ gives $o(G+X, \tilde p) = \R$, which  contradicts  (\ref{eq1}). Thus $o(G, \rwhat 0) = \L$.
\end{proof}
\begin{cor}\label{cor:Gle0}
    If the bidding game $G<0$ then, for any budget partition $\tilde p\in \B$,  $o(G, \tilde p) = \R$.
\end{cor}
\begin{proof}
    By symmetry, this follows from Theorem~\ref{thm:Ggr0}.
\end{proof}

We demonstrated the influence of ``inverse'' in game comparison. Sometimes, we can ensure the existence of an inverse. Towards this end, we next study number games.

\section{Numbers}\label{sec:numbers}

Recall that a game form $G$ is a {\em number} if for all $G^L$ and for all $G^R$, $G^L<G<G^R$, and all options are numbers. In alternating play, numbers are a subgroup of all game forms. In this section, we prove that the same is true in the bidding set up.

\begin{thm}[Numbers are Invertible]\label{thm:numberinv} Consider a game $G$.
If $G$ is a number, then $G+\bar G = 0$.
\end{thm}
\begin{proof}
By Corollary~\ref{cor:main} we must show:
\begin{enumerate}
    \item $o(G + \bar G, 0) = \L$, with a Left 0-bid strategy and
    \item $o(G + \bar G, \rwhat{\tb}) = \R$, with a Right 0-bid strategy.
\end{enumerate}
By symmetry, it suffices to prove item (1).

Consider the game $(G+ \bar G, 0)$. Left starts by bidding $0$. Right will win this bid, so
 without loss of generality suppose he also bids $0$ (bidding $1$ to keep the marker cannot benefit him because of Marker Worth Lemma). Right plays to either $(G^R + \bar G, \rwhat 0)$ or $(G + {\bar{G^L}}, \rwhat 0)$, and we must prove that Left wins both cases.\\

\noindent{\bf Case 1:} In  the game $(G^R + \bar G, \rwhat 0)$ she will again bid $0$, and if she wins the bid, she plays to  $(G^R + {\bar{G^R}}, 0)$. The outcome is $L$ by induction. 

   Further, in the game $(G^R + \bar G, \rwhat 0)$,  if Right outbids Left then he can play to either  $(G^{RR} + \bar G, \rwhat 1)$ or $(G^R + {\bar{G^L}}, \rwhat 1)$. Since $G^R$ is a number, we have $G^R<G^{RR}$. Thus $o(G^R + \bar G, \rwhat 1) \le o(G^{RR} + \bar G, \rwhat 1)$. By induction $o(G^R + \bar G, \rwhat 1) = \L$, and therefore $o(G^{RR} + \bar G, \rwhat 1) = \L $.  Similarly, since $\bar{G}$ is a number, we have $\bar{G} < {\bar{G^L}}$. Thus $o(G^R + \bar{G}, \rwhat 1) \le o(G^R +{\bar{G^L}}, \rwhat 1)$.  By induction $o(G^R + \bar G, \rwhat 1) = \L$, and therefore $o(G^R + {\bar{G^L}}, \rwhat 1) = \L $. 

\noindent{\bf Case 2:} Similar to the previous case, we get  $o(G + {\bar{G^L}}, \rwhat 0) = \L$.
\end{proof}

\begin{thm}[Uniqueness of Number Inverse]\label{thm:numberuniqinv}
For any number game $G$, the unique inverse is $\bar{G}$. 
\end{thm}
\begin{proof}
Consider a number G. Let apart from $\bar{G}$, $G'$ is another inverse of G. 
We need to show,
\begin{align}
    &&\bar G &= G' && \nonumber \\
    \rm{i.e.} && o(\bar G + Y, \tilde p) &= o(G' + Y, \tilde p)~~&&\forall ~\rm{game}~Y, ~\rm{and}~\forall \tilde{p} \in \B \label{uoi}
\end{align}
Since $\bar G$ is inverse of $G$,  we have $G+\bar{G} = 0$. Then 
\begin{align}
    && o(G+ \bar{G} + X, \tilde p) &= o(X, \tilde p)~~&&\forall ~\rm{game}~X, ~\rm{and}~\forall \tilde{p} \in \B \label{uoi2}
\end{align}
For any game Y in $\eqref{uoi}$, taking $X= G'+Y$ in $\eqref{uoi2}$ gives the result. 
\end{proof}

 \begin{thm}[Numbers are Additive]\label{thm:numadd}
 For any total budget, if $G$ and $H$ are numbers, then $G+H$ is a number. 
 \end{thm}
 \begin{proof}
 Study the Left option $G^L+H$. Since $G^L<G$, then, by Theorems~\ref{thm:strictorder} and \ref{thm:numberinv}, $G^L+H<G+H$.
 \end{proof}
Thus, the submonoid of all games that are numbers is a group.


\section{Integer games}\label{sec:integer}
In alternating play, integer games is a subgroup of games that are numbers. We will prove that this continues to hold when $\tb>0$.

Can Right ever win the game $1:=\cg{0}{\varnothing}$? For Right to win, Left must win two consecutive bids. This is not possible in optimal play since Left can bid 0 twice, and either lose the first or second bid depending on who starts with the marker.  
\begin{obs}\label{obs:Rcannot}
Given any total budget and any budget partition $\tilde p\in \B$, Left wins the game $1=\cg{0}{\varnothing}$. Notice that she wins by using a $0$-bid strategy.
\end{obs}

In fact, the game form $1>0$. 
For all $n\in \mathbb N$, define the {\em positive integer} game form $n$ by $n=\cg{n-1}{\varnothing}$. 
\begin{prop}\label{prop:Lwinn}
Consider any $\tb$ and any budget partition $\tilde p\in \B$. For  any $n \in \mathbb{N}$, Left wins the game form $n$ with a 0-bid strategy. 
\end{prop}
\begin{proof}
    From Monotonicity it is sufficient to prove the following with Left 0-bid strategy:
    \begin{enumerate}[i)]
        \item $o(n, 0) = \L$ and
        \item $o(n,\rwhat{0}) = \L$.
    \end{enumerate}
    \textit{Proof of i). }Consider the game $(n,0)$. Right wins the first bid, but does not have any move to play. Thus Left wins with her 0-bid strategy.\\
    \noindent \textit{Proof of ii). } Consider the game $(n, \rwhat{0})$. If Right outbids Left then he immediately loses, as he cannot play. Thus assume that Left wins the first bid and plays to $(n-1, 0)$. Left wins using part i) or by playing the last move.  
\end{proof}
\begin{obs}\label{obs:Rwinnbar}
Consider any $n \in \mathbb{N}$.  A symmetric result to Proposition~\ref{prop:Lwinn} gives that, for any $\tb$ and any budget partition $\tilde p\in \B$, Right wins the game form $\bar n$ with a 0-bid strategy.     
\end{obs}
\begin{obs}\label{obs:ng0nbarl0}
    From Proposition~\ref{prop:Lwinn},  Observation ~\ref{obs:Rwinnbar} and Corollary~\ref{cor:main}, we have: for any $n \in \mathbb{N},$ $ n>0$ and $ \bar n <0$.
\end{obs}
Using Observation~\ref{obs:ng0nbarl0} in the proof of Theorem~\ref{thm:noninv}, shows that {\em switches}, i.e. games such as  $\cg{1}{-1}$, do not have inverses.




\begin{thm}\label{thm:integersinv}
Consider $n\in \mathbb N$. Then the disjunctive sum $ n+\bar { n} =0$.
\end{thm}
\begin{proof}
Similar to the proof of Theorem \ref{thm:numberinv}, it is sufficient to show $o(n+ \bar n, 0) = \L$ with a Left 0-bid strategy. 
Consider the game $(n+ \bar n, 0)$. By Left's $0$-bid strategy, Right will win the first bid, so (by MMW) suppose that he also bids $0$. Right plays to $(n + \bar{n-1}, \rwhat 0)$. In this position, Left bids $0$ again and if she wins the bid, she plays to $(n-1+ \bar{n-1}, 0)$. Here Left wins  by induction. However if Right outbids Left, then he must play to $(n+\bar{n-2}, \rwhat 1)$ (if $n=1$, then Right immediately loses). If Left wins the next bid, she plays to $(n-1 + \bar{n-2}, 1)$. She wins this using induction and Monotonicity. If Right wins the bid, he plays to $(n+ \bar{n-3}, \rwhat 2)$ (if possible, and otherwise loses immediately). 

 Since both $n$ and $\tb$ are finite, 
 he cannot continue winning the bid indefinitely while having options to play. Therefore, he will eventually end up with no move or he loses the bid, and when this happens, Left wins with a 0-bid strategy by induction and  Monotonicity.
\end{proof}
Observe that from Theorem~\ref{thm:integersinv}, the inverse of $n$ is $\bar n$.  The set $\{n,\bar n: n\in \nz \}$ is the set of all {\em integer game} forms. The set $\{\bar n: n \in \mathbb N\}$ is the set of all {\em negative integer} game forms. Hence note that the inverse of a positive integer game form is a negative integer game form and vice versa. In the coming, we use interchangeably $-n = \bar n$. Also note that $-n = \cg{\varnothing}{-n+1}$.
\begin{thm}\label{thm:integersadd}
Suppose that $(n)$ and $(m)$ are integer game forms. Then the disjunctive sum $(n)+(m)$ is an integer game form that equals the integer game form $(n+m)$.\footnote{Here, and in Theorem~\ref{thm:integertotalorder}, it is convenient to  use the notation  $(n)$ for the game form $n$. 
}
\end{thm}
\begin{proof}
If any of the integers $n$ or $m$ equals $0$, then by Theorem~\ref{thm:zero}, there is nothing to prove. 

\noindent \textbf{Case i)} Let us first assume $n,m>0$, that is $(n)=\cg{(n-1)}{\nil}$ and $(m)=\cg{(m-1)}{\nil}$. Right cannot move in the sum $(n)+(m)$. Left moves to $(n-1)+(m)$ or $(n)+(m-1)$. Let $(k)=\cg{(n+m-1)}{\nil}$. By induction, regarded as games, $(n+m-1)=(n-1)+(m)$ and $(n+m-1)=(n)+(m-1)$. Hence, Left has the same options in $(k)$ as in $(n)+(m)$. Hence the games are the same.

 \noindent \textbf{Case ii)} The case when both $n,m<0$ is symmetric. 

\noindent \textbf{Case iii)} Assume $n>0$ and $m<0$. 

\noindent \textbf{Subase i)} \textbf{$n< -m$} 
\begin{align*}
    (n)+(m) &= (n) + (m+n-n) &&\\
    &= (n) + (- n) + (m+ n)&& ;(-n)<0 ~, ~(m+n)<0\\
    &= (m+ n)&&; \text{By using Theorem~\ref{thm:integersinv}}
\end{align*}
\noindent \textbf{Subcase ii)} \textbf{$n>- m$} 
\begin{align*}
    (n)+(m) &= (n+m-m) + (m) &&\\
    &=(n+m) + (-m) + (m)&& ;-m>0 ~, ~(n+ m)>0\\
    &= (n+ m)&&
\end{align*}
\noindent \textbf{Subcase iii)} \textbf{$n=- m$} \\
The equality $(n)+(m)=(n+m)$ follows by  Theorem~\ref{thm:integersinv}.




\end{proof}

\begin{thm}\label{thm:integertotalorder}
Consider any total budget $\tb$. If the integers $n>m$, then the game forms $(n)>(m)$. 
\end{thm}
\begin{proof}
     By using Theorems~\ref{thm:strictorder} and \ref{thm:integersinv}, it suffices to prove that $(n + \bar{m})>0$. According to Corollary~\ref{cor:main}, we prove the following:  
    \begin{enumerate}[i)]
        \item $o((n +\bar m), 0) = \L$ with Left 0-bid strategy and
        \item $o((n+\bar m), \rwhat{\tb}) = \L$.
    \end{enumerate}
    \textbf{Case 1. }Consider $m>0$. \\
    \textit{Proof of i). } Consider the game $((n+ \bar m), 0)$. Right wins the first bid, so (by MMW) he plays to $((n + \bar{m-1}), \rwhat 0)$. Now Left  again bids $0$; if she wins this bid then she plays to $((n-1 +\bar{m-1}), 0)$. Left wins this by induction. However if Right outbids her then he plays to $((n+ \bar{m-2}), \rwhat{1})$ (if $m=1$, then Right immediately loses, however if $m>1$ then only this case arises). Left will again bid $0$. If she wins the bid, she plays to $((n-1 + \bar{m-2}), 1)$. She wins this using induction and Monotonicity. If Right wins the bid, he plays to $((n + \bar{m-3}), \rwhat{2})$ (if possible, and otherwise loses immediately). 
    
    Since both $m$ and $\tb$ are finite, 
 he cannot continue winning the bid indefinitely while having options to play. Therefore, he will eventually end up with no move or she loses the bid, and when this happens, Right wins with a 0-bid strategy by induction and  Corollary~\ref{cor:Rwinsumdyad}. 
    
    Since $m$ is  finite, he cannot keep winning the bid indefinitely and have option to play. Thus either Right will lose by winning the bid and have no option to play or he loses the bid and when this happens, Left wins by induction and Monotonicity.\\  
\noindent \textit{Proof of ii). } Left wins the first bid and plays to $((n-1+ \bar{m}), \tb)$. She will again bid $0$, which forces Right to play. He moves to $((n-1 + \bar{m-1}) ,\rwhat{\tb})$. Thus Left wins by induction. \\
\textbf{Case 2. } In case $n=0$ or $m=0$,   the result follows from Observation \ref{obs:ng0nbarl0}.\\
\textbf{Case 3.} Consider $n>0$ but $m<0$. Then $(n)>0$ and $(\bar m )>0$. Thus from Theorem~\ref{thm:integersadd} we have $(n+\bar m)>0$.\\
\textbf{Case 4. } Consider $n<0$. This case is symmetric to Case~1, so by analogy  $(n+\bar m)>0$.
\end{proof}

From Theorem~\ref{thm:integertotalorder}, we have that the  integer game forms follow the natural total order.
We conclude this part by showing that, for a given total budget, the integer game forms form a subgroup of all bidding games. First, by using previous results in this section, we demonstrate a ``simplicity'' lemma for integers.

\begin{lem}\label{lem:intsimpl}
For all integers $n$, $n=\cg{n-1}{n+1}$.
\end{lem}
\begin{proof}
Since the inverse of $n$ is $-n$, it suffices to demonstrate  that $\cg{n-1}{n+1}-n=0$. \\
\textbf{Case i). }Consider $n>0$. By Corollary~ \ref{cor:main}, we must prove the following:
\begin{enumerate}[i)]
    \item $o(\cg{n-1}{n+1} -n, \rwhat{\tb}) = \R$ with Right 0-bid strategy.
    \item $o(\cg{n-1}{n+1} -n, 0) = \L$ with Left 0-bid strategy.
\end{enumerate}
\textrm{Proof of i). } Consider the game $(\cg{n-1}{n+1} -n, \rwhat{\tb})$. Left wins the first bid and plays to $(n-1-n, \tb)$. From Theorems~\ref{thm:integersinv} and \ref{thm:integersadd} it is equivalent to $(-1, \tb)$. Right wins this using 0-bid strategy from Observation~\ref{obs:Rwinnbar}. \vspace{0.2cm}

\noindent\textrm{Proof of ii). } Consider the game $(\cg{n-1}{n+1} -n, 0)$. By Left's 0-bid strategy, Right will win the first bid. He plays to either $(n+1-n, \rwhat 0)$ or $(\cg{n-1}{n+1} -n+1, \rwhat{0})$. 

From Theorem~\ref{thm:integersinv} and \ref{thm:integersadd} $(n+1-n, \rwhat 0)$ is equivalent to $(1, \rwhat 0)$. Left wins this using a 0-bid strategy from Proposition~\ref{prop:Lwinn}. Hence playing in first component is losing for Right. So, we may assume Right will not play in the first component if similar situation arises. 

Next, in the game $(\cg{n-1}{n+1} -n+1, \rwhat{0})$,  Left will again bid 0. If she wins the bid then she plays to $(n-1-n+1, 0)$. From Theorem~\ref{thm:integersinv} and \ref{thm:integersadd} it is equivalent to $(0, 0)$, which Left wins because she made the last move. If he outbids her, then he plays to $(\cg{n-1}{n+1}-n+2, \rwhat 1)$, because of the reason in the previous paragraph. Since both $n$ and $\tb$ are finite,  he cannot continue winning the bid indefinitely while having options to play. Therefore, he will eventually end up with no move or he loses the bid, and when this happens, Left wins with a 0-bid strategy by Theorem~\ref{thm:integersinv}, \ref{thm:integersadd} and Proposition~\ref{prop:Lwinn}. \vspace{0.2cm}
 
\noindent \textbf{Case ii). } Consider $n=0$. By Corollary~ \ref{cor:main}, we must prove the following:
\begin{enumerate}[i)]
    \item $o(\cg{-1}{1}, \rwhat{\tb}) = \R$ with Right 0-bid strategy.
    \item $o(\cg{-1}{1}, 0) = \L$ with Left 0-bid strategy.
\end{enumerate}
In the game $(\cg{-1}{1}, \rwhat{\tb})$, Left will win the first bid and plays to $(-1,\tb)$. This Right wins from Observation~\ref{obs:Rwinnbar} with his 0-bid strategy. Symmetrically Left wins $(\cg{-1}{1}, 0)$ with her 0-bid strategy. \vspace{0.2cm}
 
\noindent \textbf{Case iii). }Consider $n<0$. This case is symmetric to Case i, hence have the same result.
\end{proof}
\begin{cor}\label{cor:integersnumbers}
    Integers are numbers.
\end{cor}
\begin{proof}
It follows from Lemma~\ref{lem:intsimpl} and Theorem~\ref{thm:integertotalorder}.
\end{proof}


\begin{thm}\label{thm:intsubgroup}
The set of all integer game forms is a subgroup of the numbers.
\end{thm}
\begin{proof}
It follows from Corollary~\ref{cor:integersnumbers}, Theorems~\ref{thm:integersinv} and  \ref{thm:integersadd} .
\end{proof}

Perhaps the most central remaining problem of this section is as follows. 

\begin{problem}
Suppose that the birthday of $G$ is smaller than the integer $n>0$. Is it true that $n>G$? Does Right require an optimal 0-bid strategy in the game $G-n$? 
\end{problem}


\section{Dyadic rational games}\label{sec:dyad}
The bidding integers are thus well understood. The obvious next question concerns bidding dyadic rationals. The first few questions would be:

\begin{problem}
Is the bidding game $1/2=\cg{0}{1}>0$?
\end{problem}

\begin{problem}
Is the bidding game $1/2+ \bar{1/2}=0$?
\end{problem}

\begin{problem}\label{prob:halfhalfone}
Is the bidding game $1/2+ 1/2=1$?
\end{problem}
These types of problems will be resolved in a general form. Let the game $1/2^0$ be the game $1$. 
For $k\in \mathbb N$, we define the positive dyadic rational game forms as $1/2^k=\cg{0}{1/2^{k-1}}$. For $n\in \nz$, $n/2^k$ is the disjunctive sum of $n$ copies of $1/2^k$.

\begin{prop}\label{prop:Lwindyad}
Consider any $\tb$ and any budget partition $\tilde p\in \B$. For  any $k \in \mathbb{N}$, Left wins the game  $(1/2^k,\tilde p)$ with a 0-bid strategy. 
\end{prop}
\begin{proof}
    By Monotonicity it is sufficient to prove that, with a Left 0-bid strategy:
    \begin{enumerate}[i)]
        \item $o(1/2^k,\rwhat{0}) = \L$, and
        \item $o(1/2^k, 0) = \L$.
        \end{enumerate}
\noindent \textit{Proof of i).} Consider the game $(1/2^k,\rwhat{0})$ together with a Left 0-bid strategy. If she wins the bid, then she will make the last move and win. If Right outbids her, he must play to $(1/2^{k-1},\rwhat{1})$, which she wins by induction.

\vspace{0.2cm}
\noindent \textit{Proof of ii).} Consider the game $(1/2^k, 0)$ together with a Left 0-bid strategy. Right wins the first bid and moves to $(1/2^{k-1},\rwhat{0})$ (by MMW). Left wins this from the first part of the proof.
\end{proof}

\begin{cor}\label{cor:Rwindyad}
    Consider any $\tb$ and any budget partition $\tilde p\in \B$. For  any $k \in \mathbb{N}$,  Right wins $(\bar{1/2^{k}},\tilde p)$ with a 0-bid strategy. 
\end{cor}
\begin{proof}
    By symmetry, this follows from Proposition~\ref{prop:Lwindyad}.
\end{proof}
\begin{cor}\label{cor:dyadposneg}
    For  any $k \in \mathbb{N}$, $1/2^{k}>0$  and $\bar{1/2^{k}}<0$. 
\end{cor}
\begin{proof}
    Apply Corollary~\ref{cor:main} to Proposition~\ref{prop:Lwindyad} and Corollary~\ref{cor:Rwindyad}.
\end{proof}

\begin{thm}\label{thm:Lwinsumdyad}
For all integers $k'> k\ge 0$ and for any budget partition $\tilde p \in \B$, Left wins $(1/2^k+\bar{1/2^{k'}}, \tilde p)$ with a 0-bid strategy. 
\end{thm}
\begin{proof}
    Consider integers $k'> k\ge 0$. From monotonicity it is sufficient to show
    \begin{enumerate}[i)]
        \item $o(1/2^k + \bar{1/2^{k'}}, 0) = \L$ with Left's 0-bid strategy, and 
        \item $o(1/2^k + \bar{1/2^{k'}}, \rwhat 0) = \L$ with Left's 0-bid strategy.
    \end{enumerate}
    
    Consider $k=0$. Right wins the first bid in the game $(1 + \bar{1/2^{k'}}, 0) $, and, by MMW, plays to $(1,\rwhat 0)$. By  Proposition~\ref{prop:Lwinn}, Left wins this game using a 0-bid strategy.
    
    Next, Left bids $0$ in the game $(1 + \bar{1/2^{k'}}, \rwhat{0})$. If she wins the bid then she wins the game with a 0-bid strategy using induction by playing to $(1 + \bar{1/2^{k'-1}}, 0)$. However if Right outbids her, then by MMW, he plays to $(1, \rwhat 1)$. By Proposition~\ref{prop:Lwinn}, Left wins this with a 0-bid strategy.
    
    Consider $k>0$. Right wins the first bid in the game $(1/2^k + \bar{1/2^{k'}}, 0)$,  and, by MMW, plays to either $(1/2^k, \rwhat 0)$ or $(1/2^{k-1} + \bar{1/2^{k'}}, \rwhat 0)$. By Proposition~\ref{prop:Lwindyad}, Left wins $(1/2^k, \rwhat{0})$ with a 0-bid strategy. Hence playing in the second component is losing for Right. Therefore we may assume that Right will not play in the second component if a similar situation arise.  In the other game $(1/2^{k-1} + \bar{1/2^{k'}}, \rwhat 0)$, Left continues to use her 0-bid strategy. If she wins the bid, then she plays to $(1/2^{k-1} + \bar{1/2^{k'-1}},0)$ and wins with 0-bid strategy by induction. However, if Right outbids her, then by MMW, he plays to  $(1/2^{k-2} + \bar{1/2^{k'}}, \rwhat 1)$ (if $k=1$, he loses immediately). Since both $k$ and $\tb$ are finite, he cannot continue winning the bid indefinitely while having options to play. Therefore, he will eventually end up with no move or he loses the bid, and when this happens, Left wins by induction.

    Next, in the game $(1/2^k + \bar{1/2^{k'}}, \rwhat{0})$, by Left's 0-bid strategy, if she wins the bid, then she plays to $(1/2^k + \bar{1/2^{k'-1}}, 0)$. Left bids $0$ again. Right wins the bid and, by MMW, plays to $(1/2^{k-1} + \bar{1/2^{k'-1}}, \rwhat 0)$. Left wins this using induction. However, in the game $(1/2^k + \bar{1/2^{k'}}, \rwhat{0})$, if Right outbids her, then by MMW, he plays to either $(1/2^{k}, \rwhat{1})$ or  to $(1/2^{k-1} + \bar{1/2^{k'}}, \rwhat{1})$.  From Proposition~\ref{prop:Lwindyad}
 and from previous paragraph Left wins both using her 0-bid strategy.
 \end{proof}
 
\begin{cor}\label{cor:Rwinsumdyad}
    For all integers $k'> k\ge 0$ and for any budget partition $\tilde p \in \B$, Right wins $(1/2^{k'}+\bar{1/2^{k}}, \tilde p)$ with a 0-bid strategy. 
\end{cor}
\begin{proof}
    By symmetry, this follows from Theorem~\ref{thm:Lwinsumdyad}.
\end{proof}
\begin{cor}\label{cor:dyadsumineq}
For all integers $k'> k\ge 0$, the disjunctive sum  $1/2^k+\bar{1/2^{k'}}>0$ and $1/2^{k'}+\bar{1/2^{k}}<0$.  
\end{cor}

\begin{proof}
    This follows by applying Corollary~\ref{cor:main} to Theorem~\ref{thm:Lwinsumdyad} and Corollary~\ref{cor:Rwinsumdyad}.
\end{proof}
We prove that dyadic rational games have their conjugates as inverses.

\begin{thm}\label{thm:dyadinv}
    Consider $k\in \mathbb N$. Then the disjunctive sum $ 1/2^{k}+\bar{1/2^{k}} =0$.
\end{thm}
\begin{proof}
    Similar to the proof of Theorem \ref{thm:numberinv}, it is sufficient to show $o(1/2^{k}+\bar{1/2^{k}}, 0) = \L$ with a Left 0-bid strategy. Consider the game $(1/2^{k}+\bar{1/2^{k}}, 0)$. By Left's $0$-bid strategy, Right wins the first bid. By MMW, he plays to either $(1/2^{k-1} + \bar{1/2^k}, \rwhat{0})$ or $(1/2^k, \rwhat{0})$. By Proposition~\ref{prop:Lwindyad}, Left wins $(1/2^k, \rwhat{0})$ with a 0-bid strategy. Thus, playing in the second component is losing for Right. Therefore we may assume that Right will not play in the second component when a similar situation arise. 
    
    Next, in the game  $(1/2^{k-1} + \bar{1/2^k}, \rwhat{0})$, Left will again bid $0$. If she wins the bid she plays to $(1/2^{k-1}+ \bar{1/2^{k-1}}, 0)$ and wins by induction using a 0-bid strategy. However, if Right outbids her, then he plays to $(1/2^{k-2} + \bar{1/2^k}, \rwhat{1})$, because of the reason in the previous paragraph (if $k=1$ he loses immediately). Since both $k$ and $\tb$ are finite, he cannot continue winning the bid indefinitely while having options to play. Therefore, he will eventually end up with no move or he loses the bid, and when this happens, he wins with a 0-bid strategy by Proposition~\ref{prop:Lwindyad} or Theorem~\ref{thm:Lwinsumdyad}.
\end{proof}

\begin{cor}\label{cor:dyadinv}
    Consider $n,k\in \mathbb N$. Then the disjunctive sum $ n/2^{k}+\bar{n/2^{k}} =0$.
\end{cor}
\begin{proof}
    This follows from Theorem~\ref{thm:dyadinv}. 
\end{proof}
By Corollary~\ref{cor:dyadinv}, the inverse of $n/2^{k}$ is $\bar{n/2^{k}}$ and vice versa.  The set $\mathbb{D} = \{n/2^{k}: n\in \mathbb{Z}, k \in \mathbb{N} \}$, represents the set of all {\em dyadic rational game} forms. The set $\{\bar{n/2^{k}}: n\in \mathbb{N}, k \in \mathbb{N} \}$, constitutes the set of all {\em negative dyadic rational game} forms. In the coming, we use interchangeably $-n/2^{k} = \bar{n/2^{k}}$. 

\begin{thm}\label{thm:dyadgeneral}
    For any $n,k \in \mathbb{N}$, $n/2^{k}>0$ and $-n/2^{k}<0$. 
\end{thm}
\begin{proof}
By Corollary~\ref{cor:dyadposneg}, for any $k \in \mathbb{N}$, $1/2^k>0$. Since, by Theorems~\ref{thm:dyadinv}, $1/2^k$ is invertible, therefore, by Theorems~\ref{thm:strictorder}, we get $2/2^k >1/2^k$. Consequently, by Theorems~\ref{thm:strictordertransitivity}, we get $2/2^k >0$. By recursively applying the above steps, for any $n \in \mathbb{N}$,  we get $n/2^k >0$.

Furthermore, since $-n/2^{k}$ is inverse of $n/2^{k}$ and    $n/2^{k}> 0$, therefore,  by Corollary~\ref{cor:inverseineq}  it follows that $-n/2^{k}<0$.
\end{proof}

 \begin{lem}\label{lem:dyadnum}
 For any $k\in \mathbb N$, $1/2^k= \cg{0}{1/2^{k-1}}$ is a number.
 \end{lem}
 \begin{proof}
By Corollary~\ref{cor:dyadsumineq}, $1/2^k -1/2^{k-1}<0$. Since $1/2^k$ is invertible, therefore, by Theorem~\ref{thm:strictorder}, $1/2^k <1/2^{k-1}$. Further, by Corollary~\ref{cor:dyadposneg}, $1/2^k> 0$. Hence, $1/2^k$ is a number.
 \end{proof}
We  now show that the set of dyadic rational game forms is closed under disjunctive sum.
 \begin{lem}\label{lem:adddyad}
 For any $k\in \mathbb N$, the disjunctive sum  $1/2^k + 1/2^k = 1/2^{k-1}$.
 \end{lem}
 \begin{proof}
 Consider $k\in \mathbb N$. By Theorem~\ref{thm:orderpres} and~\ref{thm:dyadinv}, we show $1/2^k + 1/2^k - 1/2^{k-1} =0$. Further, by Corollary~\ref{cor:main}, it is sufficient to show
    \begin{enumerate}[i)]
         \item $o(1/2^k+1/2^k-1/2^{k-1}, 0)= \L$ with a Left's 0-bid strategy, and
         \item $o(1/2^k+1/2^k-1/2^{k-1}, \rwhat{\tb})= \R$ with a Right 0-bid strategy.
     \end{enumerate}
\noindent \textrm{Proof of i).} Consider the game $(1/2^k+1/2^k-1/2^{k-1}, 0)$. By Left's 0-bid strategy, Right wins the first bid.  By MMW, he plays to either $(1/2^{k-1} + 1/2^k-1/2^{k-1}, \rwhat 0)$ or $(1/2^k+1/2^k, \rwhat 0)$.  By Theorem~\ref{thm:dyadinv}, $(1/2^{k-1} + 1/2^k-1/2^{k-1}, \rwhat 0)$ is the same as $(1/2^k, \rwhat 0)$, which Left wins with a 0-bid strategy (see Proposition~\ref{prop:Lwindyad}). By Theorems~\ref{thm:dyadgeneral}, $1/2^k+1/2^k>0$.  Thus, by Theorems~\ref{thm:Ggr0}, Left wins  $(2/2^k,\rwhat 0)$. Further, by Lemma~\ref{lem:dyadnum}, $1/2^k$ is a number and number games form a group, thus, $2/2^k$ is a number. Therefore, Proposition~\ref{prop:0bidnumber} ensures an optimal 0-bid strategy of Left in $(2/2^k,\rwhat 0)$.\\

\noindent \textrm{Proof of ii).} Consider the game $(1/2^k+1/2^k-1/2^{k-1}, \rwhat{\tb})$. We must prove that Right wins this game with a 0-bid strategy. By Right's 0-bid strategy, Left wins the first bid. By MMW, she plays to either $(1/2^k-1/2^{k-1}, \tb)$ or  $(1/2^k + 1/2^k - 1/2^{k-2}, \tb)$ (if $ k =1$, she only gets to play in the first two components). 

By Corollary~\ref{cor:Rwinsumdyad}, Right wins $(1/2^k-1/2^{k-1}, \tb)$ using a 0-bid strategy. Thus, playing in the first two components are losing for Left. Therefore we may assume that she will not play in those components if similar situations arise. 

For $(1/2^k + 1/2^k - 1/2^{k-2}, \tb)$,
 if Right wins the bid by his 0-bid strategy, then he plays to $(1/2^k + 1/2^{k-1} - 1/2^{k-2}, \rwhat \tb)$. Using induction, it is equivalent to $(1/2^k + 1/2^{k-1} - 1/2^{k-1} -1/2^{k-1}, \rwhat \tb)$, which, by Theorem~\ref{thm:dyadinv}, is equal to $(1/2^k - 1/2^{k-1}, \rwhat \tb)$. By Corollary~\ref{cor:Rwinsumdyad}, Right wins  using a 0-bid strategy. However, if Left outbids him in the game $(1/2^k + 1/2^k - 1/2^{k-2}, \tb)$, then she plays to $(1/2^k + 1/2^k - 1/2^{k-3}, \tb-1)$, because of the reason in the previous paragraph (if $ k=2$ she loses immediately).
 Since both $k$ and $\tb$ are finite, 
 she cannot continue winning the bid indefinitely while having options to play. Therefore, she will eventually end up with no move or she loses the bid, and when this happens, Right wins with a 0-bid strategy by induction and  Corollary~\ref{cor:Rwinsumdyad}.  
 \end{proof}
\begin{thm}\label{thm:adddyad}
 Let $n$ and $m$ be odd integers. For any total budget, the disjunctive sum $n/2^k$ + $m/2^{k-j}$ is a dyadic rational, namely $(n+2^{j}m)/2^k$.
\end{thm}
\begin{proof}
By definition, $m/2^{k-j}$ is the disjunctive sum of $m$ copies of $1/2^{k-j}$ and by Lemma~\ref{lem:adddyad}, $1/2^{k-j}$ is the disjunctive sum of  $1/2^{k-j+1}$ and $1/2^{k-j+1}$. Consequently, $m/2^{k-j}$ is the disjunctive sum of $2m$ copies of $1/2^{k-j+1}$. Proceeding in a similar way, we get $m/2^{k-j}$ is the disjunctive sum of $2^jm$ copies of $1/2^{k}$. Therefore the disjunctive sum of $n$ copies of $1/2^k$ and $2^jm$ copies of $1/2^{k}$ is $(n+2^jm)$ copies of $1/2^{k}$.
\end{proof}
 
\begin{lem}\label{lem:dyadsimp}
For any  $m\in \mathbb N$, $m/2^k=\cg{(m-1)/2^k}{(m+1)/2^k}$.
\end{lem}
\begin{proof}
By the definition of disjunctive sum and by using Lemma~\ref{lem:adddyad}:
\begin{align*}
     m/2^k &= \cg{(m-1)/2^k}{(m-1)/2^k + 1/2^{k-1}};\\
    &= \cg{(m-1)/2^k}{(m-1)/2^k+2/2^{k}};\\
    &= \cg{(m-1)/2^k}{(m+1)/2^k}.
\end{align*} 
\end{proof}
\begin{thm}\label{thm:dyadnumbers}
The dyadic rational game forms are numbers.
\end{thm}
\begin{proof}
 For any $m\in \mathbb{Z}$ and $k \in \mathbb{N}$, consider a dyadic rational game  $G=m/2^k$. From Lemma~\ref{lem:dyadsimp}, $ m/2^k = \cg{(m-1)/2^k}{(m+1)/2^k} $.  By Corollary~\ref{cor:dyadposneg} and Theorem~\ref{thm:dyadinv} $G^R=(m+1)/2^{k}>G$ and $G^L=(m-1)/2^k<G$. Hence, dyadic rational game forms are numbers. 
\end{proof}
\begin{cor}\label{cor:dyadsubgroup}
The set of all integer game forms is a subgroup of the numbers.
\end{cor}
\begin{proof}
This follows from Theorems~\ref{thm:dyadnumbers}, \ref{thm:adddyad} and Corollay~\ref{cor:dyadinv}.
\end{proof}

\section{Infinitesimals and 0-games}\label{sec:infinitesimal}
\noindent Having understood the numbers and its subgroups, next we look into infinitesimals. 
\begin{defi}[Infinitesimal] Consider a total budget $\tb$.
A game form $G$ is infinitesimal if for all $ k \in \mathbb N, 1/2^k<G<1/2^k$.
\end{defi}
We have seen in Section~\ref{sec:nogroup} that the game $*$ does not have an inverse if $\tb>0$, whereas in alternating play $*+*=0$. However, another interesting property of the game $*$ continues to hold, it is  infinitesimal for any $\tb\ge 0$. 


\begin{thm}\label{thm:infinitesimal}
Consider any total budget. For any $k\in\mathbb N$, $-1/2^k<*<1/2^k$.
\end{thm}
\begin{proof}
Suppose $k\in\mathbb N$. By symmetry, it is sufficient to prove $*<1/2^k$. 
By Theorem~\ref{thm:strictorder}, this is equivalent to proving $*-1/2^k <0$. To prove this, by Corollary~\ref{cor:main} we prove the following:
\begin{enumerate}[i)]
    \item $o(*-1/2^k , \rwhat{\tb}) = R$,  with a Right 0-bid strategy, and 
    \item  $o(*-1/2^k , 0) = R$.
\end{enumerate}
\noindent \textrm{Proof of i).} Consider the game $(*-1/2^k , \rwhat{\tb})$. By Right’s 0-bid strategy, Left wins the first bid. By MMW, she plays to either $(*-1/2^{k-1} , \tb)$  or to $(-1/2^{k} , \tb)$. By Corollary~\ref{cor:Rwindyad}, Right wins $-1/2^{k}$ with a 0-bid strategy. Thus, playing in the first component is losing for Left. Therefore we may assume that  Left will not play in the first component if a similar situation arises. 
 
 Next, in the game $(* - 1/2^{k-1} , \tb)$, Right will again bid $0$. If he wins the bid then he plays to  $(-1/2^{k-1} , \rwhat{\tb})$ and wins by Corollary~\ref{cor:Rwindyad}, with a 0-bid strategy. However, if Left outbids him, she plays to $(*-1/2^{k-2} , \tb-1)$, because of the reason in the previous paragraph (if $k=1$ she loses immediately). 
 
Since both $k$ and $\tb$ are finite, 
 she cannot continue winning the bid indefinitely while having options to play. Therefore, she will eventually end up with no move or she loses the bid, and when this happens, Right plays in first component and wins with a 0-bid strategy by Corollary~\ref{cor:Rwindyad}.\vspace{0.2cm}
 
 \noindent \textrm{Proof of ii).} Consider the game $(*-1/2^k , 0)$. Right wins the first bid and plays to $(-1/2^k , \rwhat{0}\,)$, which he wins by Corollary~\ref{cor:Rwindyad}. 
\end{proof}
Recall the games $\up = \cg{0}{*}$ and $\down = \cg{*}{0}$.
\begin{cor}
 The games $\up$ and $\down$ are  infinitesimals.   
\end{cor}
\begin{proof}
    This is similar to the proof of Theorem~\ref{thm:infinitesimal}.
\end{proof}


In alternating play, it holds that $\cg{*}{*}=0$. But it is not true if $\tb>0$, because a player with budget $\rwhat\tb$ can win two consecutive bids and get the last move. But, since numbers are invertible, we already have seen a huge class of 0s. 
In alternating play, of course $\up+\down\; =0$. This does not hold in strict bidding play.
\begin{obs}
Consider $\tb>0$. We demonstrate that $\down + \up\; \ne 0$.
This is obviously true if $\tb\ge 3$, since Left can win three consecutive moves if she owns all budget and the marker. 
Consider the game $\down + \up$ with $\tb = 1$. Now let's assume Left holds the marker as well as all budget. Then Left will start by bidding $0$. Since Right doesn't have any other choice than bidding $0$, Left will win the bid by giving away the marker and will play to $\down$, and the game will reach $\up +\; *$. Again Left will bid $0$. This time Right holds the marker, so he will win the bid and the game goes to either $\left (\up , \rwhat 1\right )$ or $\left (*+*, \rwhat 1 \right )$. If the game reaches $\left (\up, \rwhat 1 \right)$, then Left will again bid $0$ and since she has the marker the game will reach $(0, 1)$, which Left will win. Next, if the game will reach  $\left (*+*, \rwhat 1\right )$, then Left has sufficient budget to win two consecutive moves and take the game to $0$, with Right holding the marker. Thus $o\left(\,\down + \up , \rwhat 1 \right) = \L$, which contradicts $o\left(0, \rwhat 1 \right)=\R$. Hence it is not true that  $\down + \up\; =0$. A similar argument works for $\tb=2$.  Hence $\down$ is the inverse of $\up$ if and only if $\tb = 0$. 
\end{obs}

Next follows another infinite class of 0s.

\begin{thm}\label{thm:GH}
Suppose Right has an optimal 0-bid strategy in $G<0$ and Left has an optimal 0-bid strategy in $H>0$. Then $\cg{G}{H}=0$. 
\end{thm}
\begin{proof}
By symmetry and Corollary~\ref{cor:main}, it is sufficient to show that Left wins $(\cg{G}{H}, 0)$ with a 0-bid strategy. In the game $(\cg{G}{H}, 0)$, Right wins the first bid. By MMW, he must play to $(H, \rwhat{0})$. By assumption, Left wins this with a 0-bid strategy. 
\end{proof}

\begin{cor}\label{cor:kk}
For all positive integers $k$, $\cg{-k}{k}=0$ and $\cg{-1/2^k}{1/2^k}=0$.
\end{cor}
\begin{proof}
This follows by Theorem~\ref{thm:GH}.
\end{proof}

\section{Open problems}\label{sec:future}

 In view of the uniqueness of identity (Theorem~\ref{thm:uniqueiden}), we have the following problem.

\begin{problem}
Consider $\tb>0$. Does there exist any non-trivial identity element, for any non-invertible $(\tb, G)$?\footnote{For example in alternating play, in a restriction of  impartial mis\'ere nim, $*2+*2$ can ``act like a 0'', in that  $*2=*2+*2+*2$ \cite{S2013}.}
\end{problem}

\begin{conj}
For all total budgets,  the equivalence class of $0$s is the unique idempotent (i.e. $0+0=0$).
\end{conj}

By Theorem~\ref{thm:markerlose}, if $o(G, 0)= \L$ and Left has an optimal 0-bid strategy in $(G,0)$, then $G\geq 0$. Moreover, by Observation~\ref{obs:nozero}, for any non-zero total budget, we have the example of a game $G$ such that $o(G,0)=\L$, but Left does not have a winning 0-bid strategy.
\begin{problem}
    Is there a game $G$ such that $o(G,0)=\L$,  without a winning 0-bid strategy for Left, but where $G\not\ge 0$?
\end{problem}
Observe that if $\tb>0$, $ \up$ is confused with $0$. This raises the following problem.
\begin{problem}
Are there positive infinitesimals if $\tb>0$?
\end{problem}

In view of the difference between the  definition of a number game used in this paper  and the one used in \cite{S2013} for alternating play, we believe that they are equivalent.
\begin{conj}
    For any game $G$, the following are equivalent:
    \begin{enumerate}[i)]
        \item $H^L < H^R$ for every subposition $H$ of $G$ and every $H^L$ and $H^R$. 
        \item $H^L < H< H^R$ for every subposition $H$ of $G$ and every $H^L$ and $H^R$.
    \end{enumerate}
\end{conj}
 We believe that each invertible game is a number. The following conjecture would be an immediate consequence (by Theorem~\ref{thm:numberinv}). 

\begin{conj}\label{conj:invconj}
If a bidding game $G$ has an inverse, $H$, then $H=\bar G$.
\end{conj}

Since defined numbers behave well in bidding play, we believe the celebrated number avoidance result for alternating play continues to hold. 
\begin{conj}[Number Avoidance/Translation]
For any total budget, if $G$ is a number and $H$ is not a number, then $G+H=\cg{G+\HL}{G+\HR}$.
\end{conj}
The next conjecture holds in alternating play.  
\begin{conj}[Number Simplicity]
For any total budget, each number is a dyadic rational.
\end{conj}
\begin{problem}\label{prob:hack}
In alternating play {\sc Hackenbush} there is no $\mathcal N$-position. What outcomes do we lose in bidding play? 
\end{problem}
\begin{conj}\label{conj:hack}
All numbers are {\sc Hackenbush} positions, and vice versa. 
\end{conj}
If this conjecture does not hold, then we ask Problem~\ref{prob:hack} for numbers.

\bibliographystyle{plain}
\bibliography{main}
\end{document}